\documentclass[12pt,epsf]{article}
\usepackage{graphicx, amsmath}

\begin{document}

\sloppy
\begin{flushright}{SIT-HEP/TM-41}
\end{flushright}
\vskip 1.5 truecm
\centerline{\large{\bf Hilltop Curvatons}}
\vskip .75 truecm
\centerline{\bf Tomohiro Matsuda\footnote{matsuda@sit.ac.jp}}
\vskip .4 truecm
\centerline {\it Laboratory of Physics, Saitama Institute of Technology,}
\centerline {\it Fusaiji, Okabe-machi, Saitama 369-0293, 
Japan}
\vskip 1. truecm
\makeatletter
\@addtoreset{equation}{section}
\def\theequation{\thesection.\arabic{equation}}
\makeatother
\vskip 1. truecm

\begin{abstract}
\hspace*{\parindent}
We study ``hilltop'' curvatons that evolve on a convex potential. 
Hilltop curvatons evolving on the Hubble-induced potential are generic
 if supergravity is assumed in the theory.
We do not consider curvatons whose potential is protected from $O(H)$ 
corrections, where $H$ is the Hubble parameter.
We assume that the effective mass of a curvaton is expressed as
 $m_\sigma = c H$, where the coefficient varies within $0.2 \le c \le 5$ 
depending on the circumstances. 
A negative mass term may lead to enhancement of curvaton fluctuation,
 which has a significant impact on the energy bound for low-scale
 inflation. 
Using a simple curvaton model and following the conventional curvaton
 hypothesis, we demonstrate the generality of this enhancement.   
\end{abstract}

\newpage
\section{Introduction}
\hspace*{\parindent}
The ``curvaton'' is an alternative formulation for generating curvature
perturbation that does not require fluctuation of the inflaton
field\cite{curvaton_1}.  
The curvaton model has several characteristic features, some of
which distinguish curvaton-induced fluctuation from inflaton fluctuation. 

In this section we briefly review the calculations related to the lower
bound of the inflation scale as derived by Lyth in
Ref.\cite{curvaton_low}.
In this paper we consider a significant relaxation of the bound. 
We initially assume that a curvaton ($\sigma$) is frozen at
$\sigma=\sigma_{osc}$ during and after inflation. 
At the beginning of curvaton oscillation, the curvaton density is given
by $\rho_\sigma \sim m_\sigma^2 \sigma_{osc}^2/2$.
Denoting the total density of all other fields by $\rho_{tot}$, the
ratio $r$ at the beginning of curvaton oscillation is 
\begin{equation}
r(t_{osc})=
\left.\frac{\rho_{\sigma}}{\rho_{tot}}\right|_{osc}
\sim \frac{m_{\sigma}^2 \sigma_{osc}^2}{H_{osc}^2 M_p^2},
\end{equation}
where $H_{osc}$ denotes the Hubble parameter at the beginning of the
oscillation. 
During the radiation-dominated epoch\footnote{This radiation could be
the radiation in a hidden sector.}, the ratio $r$ grows and
reaches 
\begin{equation}
\label{rfirst}
r \le \frac{\sqrt{H_{osc} M_p}}{T_d }
\frac{m_\sigma^2 \sigma_{osc}^2}{H_{osc}^2 M_p^2},
\end{equation}
provided the curvaton potential is dominated by a positive quadratic
term.
Here, $T_d$ is the temperature for curvaton decay. 
The basic concept of the curvaton scenario is that the curvature
perturbation 
\begin{equation}
\zeta \simeq \frac{r}{3}\frac{\delta \rho_{\sigma} }{\rho_\sigma}
\end{equation}
is generated by curvaton decay if the curvaton dominates the energy
density of the Universe. 
The spectrum of the curvature perturbation related to the curvaton
fluctuation is given by  
\begin{equation}
\label{Pzeta}
{\cal P}_{\zeta}^{\frac{1}{2}}\simeq 
\frac{2r}{3}\frac{H_I}{2\pi\sigma_{osc}},
\end{equation}
where $H_I$ denotes the Hubble parameter during inflation.
Since the observational data requires 
${\cal P}_{\zeta}^{\frac{1}{2}}=5\times 10^{-5}$, the curvaton scenario
leads to the condition 
\begin{equation}
\label{pert1}
\frac{2r}{3}\frac{H_I}{2\pi\sigma_{osc}}\simeq 5\times 10^{-5}.
\end{equation}
Using (\ref{rfirst}) and (\ref{pert1}), the condition for the
cosmological parameters related to the curvaton scenario is given by
\begin{equation}
\label{start}
\frac{2}{3}\frac{H_I}{2\pi\sigma_{osc}}
 \frac{\sqrt{H_{osc} M_p}}{T_d }
\frac{m_\sigma^2 \sigma_{osc}^2}{H_{osc}^2 M_p^2}
\ge 5\times 10^{-5}.
\end{equation}
Here, the obvious bound from nucleosynthesis is $T_d >1$ MeV,
which gives a bound
\begin{equation}
\label{constraint1}
\frac{2}{3}
\frac{H_I \sigma_{osc} m_{\sigma}^2}
{ 2\pi H_{osc}^{\frac{3}{2}} M_p ^{\frac{5}{2}}}
\ge 5\times 10^{-26}.
\end{equation}
Alternatively, we may use a lower bound for the curvaton decay rate
($\Gamma_{\sigma}\ge \frac{m_{\sigma}^3}{M_p^2}$) that leads to the
condition $T_d \simeq \sqrt{M_p \Gamma}\ge M_p
(m_{\sigma}/M_p)^{\frac{3}{2}}$.
Thus, we obtain an alternative condition 
\begin{equation}
\label{constraint2}
\frac{2}{3}
\frac{H_I \sigma_{osc} m_{\sigma}^\frac{1}{2}}
{ 2\pi H_{osc}^{\frac{3}{2}} M_p }
\ge 5\times 10^{-5}. 
\end{equation}
The four parameters ($H_I, \sigma_{osc}, H_{osc},
m_{\sigma}$) that appear in the above calculations give a cosmological
boundary condition that characterizes the model. 
Considering both (\ref{pert1}) and $r<1$, we find
\begin{equation}
\label{ineq1}
\frac{2}{3}\frac{H_I}{2\pi \sigma_{osc}}
> 5\times 10^{-5},
\end{equation}
and then from Eq.(\ref{ineq1}) and (\ref{constraint1}) we find
\begin{equation}
\label{constraint1a}
H_I > 10^{-15} \times 
\frac{H_{osc}^{\frac{3}{4}} M_p^{\frac{5}{4}}}{m_\sigma}.
\end{equation}
Alternatively, Eq.(\ref{ineq1}) and (\ref{constraint2}) give the
condition
\begin{equation}
\label{constraint2a}
H_I > 10^{-4} \times 
\frac{H_{osc}^{\frac{3}{4}} M_p^\frac{1}{2}}{m_\sigma^{\frac{1}{4}}}.
\end{equation}
In order to determine a clear bound for the inflation scale ($H_I$), we
need to fix the boundary condition. 
We assume that the curvaton mass ($m_\sigma$) is constant during the
period that we are interested in. 
Then, the oscillation of the curvaton field starts when the Hubble
parameter falls below the curvaton mass (i.e., when the Hubble parameter
reaches $H= m_\sigma$), which leads to the boundary condition 
\begin{equation}
H_{osc} \simeq m_\sigma.
\end{equation}
Since $H_I>H_{osc}$ always holds, we introduce a parameter
$p$ that measures the ratio between $H_I$ and $H_{osc}$, defined as $p^2
\equiv H_{osc}/H_{I}$. 
Using this new parameter we can replace $H_{osc}$ and $m_\sigma$ in
Eq.(\ref{constraint1a}) with $H_I$ and $p$, giving
\begin{equation}
\label{constraint1b}
H_I > M_p \times 10^{-12} \times p^{-\frac{2}{5}} 
\simeq 10^{6} p^{-\frac{2}{5}}  {\mathrm GeV}.
\end{equation}
From  Eq.(\ref{constraint2a}) we also find
\begin{equation}
\label{constraint2b}
H_I > 10^{-8} M_p p^2
\simeq 10^{10} p^2 {\mathrm GeV}.
\end{equation}
Note that the bound in Eq.(\ref{constraint2b}) is effective only when 
$p > 10^{-2}$, which suggests that a tiny $p \ll 1$ does not simply relax
the bound. 
Therefore, the above conditions lead to a lower bound for the inflation
scale; 
\begin{equation}
\label{bound-original}
H_I >10^7 {\mathrm GeV}.
\end{equation}
This result is true for the traditional curvaton that satisfies the
above boundary conditions.

To summarize, an alternative (maybe the curvaton) is needed for
low-scale inflation to avoid problems in the generation of the curvature
perturbation, but the bound in Eq.(\ref{bound-original}) suggests that
the curvaton paradigm does not accommodate a low inflation scale.
For the simplest curvaton, it is impossible to support
low-scale inflation\cite{low_inflation}, even if careful fine-tuning is
introduced to the theory.
This is a critical problem for low-scale gravity signals that might be
observed in the Large Hadron Collider(LHC).\footnote{If the fundamental
scale is as low as the TeV scale, we must introduce additional
symmetries or mechanisms to the phenomenological model so that unwanted
interactions (such as baryon-number violating interactions) do not
appear in the effective theory.
Inflation with low-scale gravity is also a problem, which may or may not
be solved by using the previous analyses\cite{low_inflation}.
However, the purpose of this paper is not to solve problems related to
low-scale gravity, though low-scale gravity is a motivation for
considering relaxation of the energy bound.}
The curvaton could be identified from the brane distance, bulk field or 
the flat direction of supersymmetric gauge theory, though we will not
specify the origin of the curvaton in this paper.
There are many model-dependent discussions for the curvaton hypothesis,
but they are beyond the scope of this paper. 
Our goal in this paper is to relax the energy condition of
Eq.(\ref{bound-original}) starting with the curvaton hypothesis so that 
the curvaton treatment can support low-scale inflation.
There are many alternatives\footnote{The alternative formulations can be
categorized as inhomogeneous reheating\cite{alt1}, inhomogeneous
preheating\cite{alt2}, $\delta N$ generation\cite{alt3} or hybrids of
these approaches\cite{alt4}. The bound for 
the curvaton cannot be applied to these alternatives.} that can be
applied to this problem.
We believe that it is worthwhile finding the simplest solution to this
problem without introducing significant modifications to the traditional
curvaton hypothesis. 
Arguments for the validity of the curvaton
hypothesis are highly model-dependent.

\subsection{Evolution on Concave potential}
\begin{figure}[ht]
 \begin{center}
\begin{picture}(550,160)(0,0)
\resizebox{15cm}{!}{\includegraphics{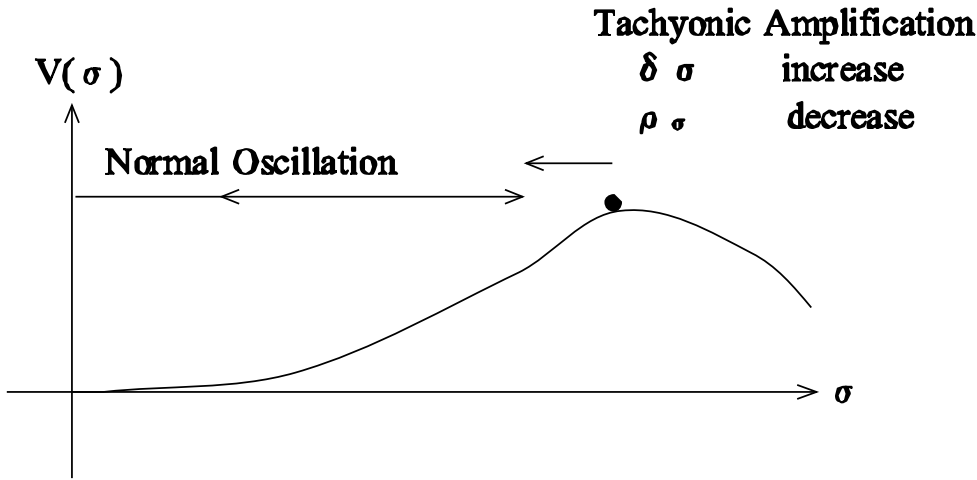}} 
\end{picture}
\caption{This figure shows tachyonic amplification of the curvaton
  fluctuation on a single-field static potential. $\delta \sigma$ grows
  near the top of the potential, but the energy density of the curvaton
  $\rho_\sigma$ decreases during this period.
  It is possible to enhance the ratio $r$ if the radiation energy
  density diminishes significantly during this period.
  This is possible if there is ``temporary rest''\cite{PQ-curvaton} of
  the curvaton near the maximum. 
  Our mechanism is different from this amplification mechanism.} 
\label{static}
 \end{center}
\end{figure}

\begin{figure}[h]
 \begin{center}
\begin{picture}(550,160)(0,0)
\resizebox{15cm}{!}{\includegraphics{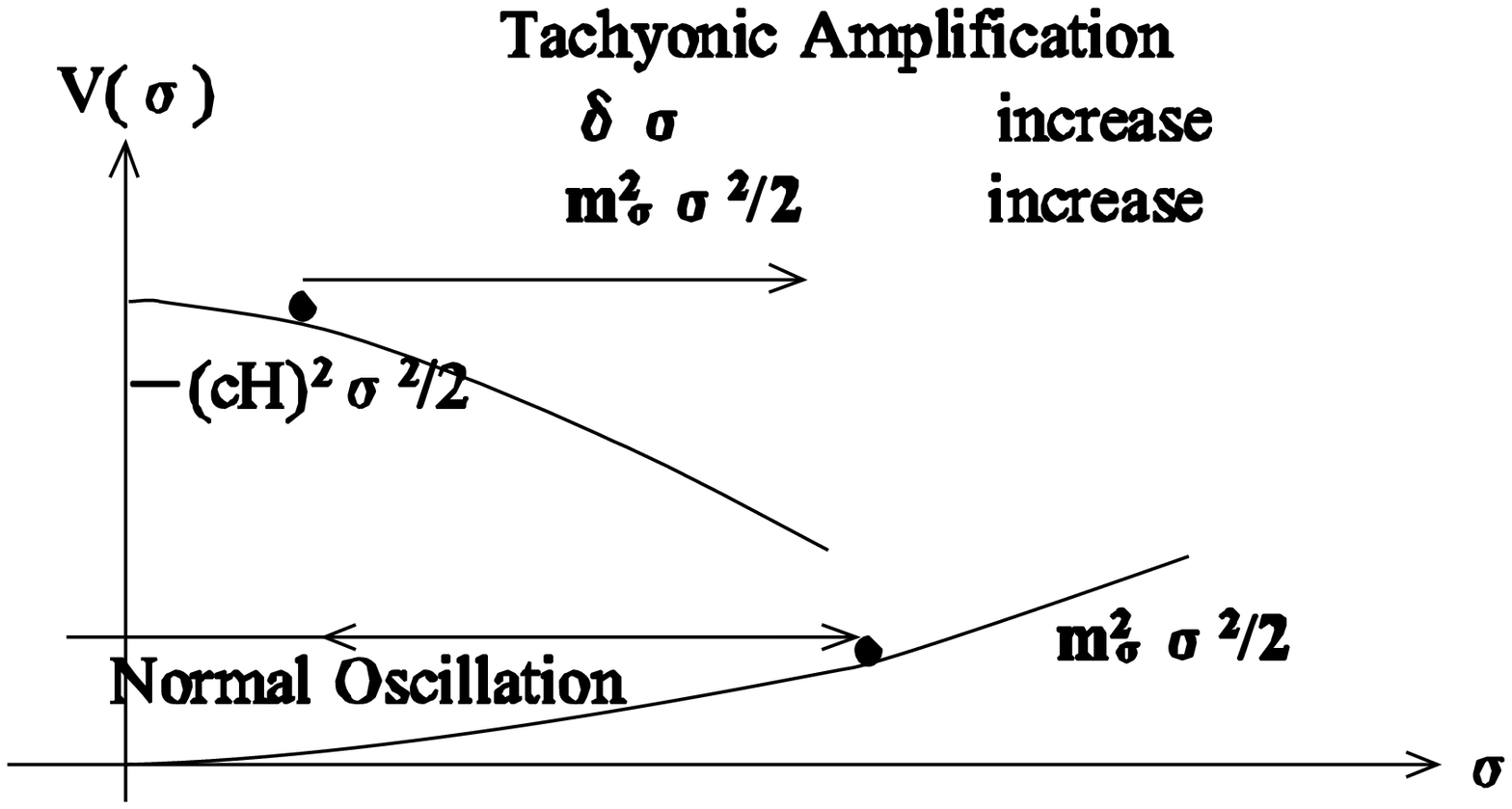}} 
\end{picture}
\caption{This figure shows tachyonic amplification of $\delta \sigma$ 
on a single-field non-static potential. Note that both $\delta \sigma$
  and $m_\sigma^2 \sigma^2/2$ increases during this period.}
\label{non-static}
 \end{center}
\end{figure}

In this section we examine the bound given in Eq.(\ref{bound-original})
if $\delta \sigma$ and $\sigma$ evolve after 
inflation but before oscillation begins.
Following the argument given in Ref.\cite{Curvaton_Dynamics}, the ratio
$\delta \sigma/\sigma$ is almost constant after inflation even if $\delta
\sigma$ and $\sigma$ evolve after inflation.
Therefore, for a concave potential $V \simeq \frac{1}{2}m_\sigma^2
\sigma^2$, we introduce a parameter $\gamma$ defined by
\begin{eqnarray}
\delta \sigma(t_{osc}) &=& \gamma \delta \sigma(t_{ini})\nonumber\\
\sigma(t_{osc}) &=& \gamma \sigma(t_{ini}),
\end{eqnarray}
where $t_{osc}$ and $t_{ini}$ denote the time when curvaton oscillation
starts and the time when inflation ends, respectively.\footnote{Note
that the radio $\delta\sigma/\sigma$ is invariant during this evolution,
and that the same result applies for the tachyonic evolution ($\gamma
>1$)  on a convex potential.}
We can use these relations to evaluate a new energy bound for $\gamma
\ne 1$.
Using $\gamma$, we find
\begin{equation}
\label{alt-constraint1a}
H_I > 10^{-15} \times 
\frac{H_{osc}^{\frac{3}{4}} M_p^{\frac{5}{4}}}{m_\sigma \gamma}
\end{equation}
rather than the expression given in (\ref{constraint1a}).
We also find
\begin{equation}
\label{alt-constraint2a}
H_I > 10^{-4} \times 
\frac{H_{osc}^{\frac{3}{4}} M_p^\frac{1}{2}}{m_\sigma^{\frac{1}{4}}\gamma},
\end{equation}
rather than (\ref{constraint2a}).
From these equations and the definition of $p$, we find 
\begin{equation}
\label{bound-alt}
H_I > \gamma^{-1} 10^7 {\mathrm GeV},
\end{equation}
which is not relaxed by curvaton evolution after inflation if $\gamma
\ll 1$. 
Therefore, for a concave potential, the bound of the energy scale is not
relaxed by curvaton evolution. 
This result supports the condition  $H_I >10^7$ GeV given in
Ref.\cite{curvaton_low}. 
It was suggested in Ref.\cite{curvaton_low} that this condition can be
relaxed by considering a ``heavy curvaton''\cite{curvaton_low} that
gives the non-standard boundary condition 
\begin{equation}
m_\sigma(t_{osc})\gg H(t_{osc}),
\end{equation}
which is due to a phase transition at the end of inflation. 
This new boundary condition significantly alters the bound and gives a
milder condition 
\begin{equation}
H_I > 10^5 {\mathrm GeV}.
\end{equation}
However, this is not sufficient to support the inflation energy scale as
low as $H_I \sim O(1)$TeV.
As we have presented in Ref.\cite{matsuda_curvaton}, if the phase
transition that gives a heavy curvaton occurs later than the primordial
inflation, it may lead to another boundary condition that may remove the
bound for $H_I$.
Other solutions for this problem have been extensively
studied\cite{alt4} using other alternative formulations. 
In this paper we consider a generic case in which both the coefficient
$c(t)$ and the sign of the quadratic term depend on the specific
conditions of the system.\footnote{See appendex A for more details.}
We show that hilltop curvatons can significantly relax the condition for
the inflation scale without introducing additional components or
symmetries to the traditional curvaton hypothesis.
Evolution of a curvaton on a convex potential that leads to tachyonic
enhancement of the fluctuation has already been discussed for a specific
model of Peccei-Quinn field curvatons\cite{PQ-curvaton}.
In this model, amplification occurs when the curvaton is stabilized for
a period near a maximum of the potential.\footnote{See also the
comment given in Ref.\cite{curvaton_low}. Another mechanism for this
enhancement has been discussed in Ref.\cite{PNGB-curvaton} 
for pseudo-Nambu-Goldstone curvatons, in which there is an additional
direction of motion other than the curvaton.} 
Although tachyonic amplification with static potential has already been
discussed for some specific models, the generality of tachyonic
enhancement is unclear for a non-static potential.
 We show that a hilltop curvaton, whose mass is not protected from modest
$O(H)$ corrections, can achieve significant enhancement of the curvaton
fluctuation.\footnote{See
Fig.(\ref{static}) and (\ref{non-static}) for the difference between
tachyonic amplification on static and non-static potentials.} 


\subsection{Hilltop curvatons}
We now discuss the enhancement of curvaton fluctuation with a convex
potential without introducing additional fine-tuning to the traditional
curvaton scenario. 
The evolution of the curvaton ($\sigma$) and its fluctuation ($\delta
\sigma$) is given in terms of the effective potential
($V(\sigma, H)$) by the equation 
\begin{equation}
\label{sigma-eq}
\ddot{\sigma}+ 3H\dot{\sigma}+V_\sigma=0,
\end{equation}
where $V_\sigma$ is the $\sigma$-derivative of the effective potential,
and by the equation
\begin{equation}
\label{delsig-eq}
\ddot{\delta \sigma} + 3H \dot{\delta \sigma} + V_{\sigma\sigma}\delta \sigma=0.
\end{equation}
Here the dot denotes differentiation with respect to the cosmic time
$t$.
Following the original curvaton hypothesis, we consider an
effective potential that represents the conventional curvaton potential
except for the explicit $O(H)$-correction;
\begin{equation}
V=\frac{1}{2}\left[m_\sigma^2 -(cH)^2\right]\sigma^2 + 
\frac{g}{n!}\frac{\sigma^n}{M_*^{n-4}},
\end{equation}
where $n$ is an integer ($n >4$), and the cut-off scale $M_*$ is
supposed to be $M_*\simeq M_p$.\footnote{In some
phenomenological models the cut-off scale $M_*$ can be much smaller
than the Planck scale.  In that case symmetries or mechanisms are
required to suppress the coefficients of higher dimensional interactions
so that the curvaton potential remains flat. We follow the original
curvaton hypothesis and do not discuss symmetries and mechanisms in this
paper since they are highly model-dependent.}
Following the curvaton hypothesis, quartic term of the curvaton
potential must be negligible.\footnote{There can be many terms that are
neglected in the above expression for the curvaton potential. 
Higher dimensional terms may have either negative or positive sign.}
The evolution of the curvaton after $cH < m_\sigma$ is precisely
the same as the conventional scenario.
Evolution before the curvaton oscillation is determined by the
$O(H)$-correction term; in our model it leads to tachyonic enhancement. 
The sign of the Hubble-induced mass term depends on the conditions and
must be negative during tachyonic enhancement. 
For simplicity, we fix the sign of the Hubble-induced mass term to be
negative, as this assumption does not reduce the generality of the
following argument. 
We assume that the expectation value of the curvaton $\sigma_0
\simeq 10^5 \times \delta \sigma_0$ is always smaller than the
effective minimum until $m_\sigma$ dominates the curvaton mass.
Note that if oscillation occurs about the effective minimum, the
curvaton fluctuation is reduced. 
Then, the evolution of the curvaton during inflation is given by 
\begin{equation}
\sigma \simeq \sigma_0 e^{F \Delta N},
\end{equation}
where 
\begin{equation}
F\equiv \sqrt{\frac{9}{4}+c^2}-\frac{3}{2}.
\end{equation}
This is the so-called fast-roll evolution\cite{fast-roll}.
Comparing (\ref{sigma-eq}) and (\ref{delsig-eq}), it is straightforward
to derive the equation for $\delta \sigma$ as 
\begin{equation}
\delta\sigma \simeq \delta\sigma_0 e^{F \Delta N}.
\end{equation}
Since the spectrum index of the curvaton perturbation is given by $n-1
\simeq 2\eta_{\sigma}$ for models with negligible slow-roll parameter
$\epsilon_H$,  we take the
condition-dependent coefficient to be $c \simeq 0.2$ during
inflation.\footnote{On the other hand, if the slow-roll parameter
$\epsilon_H$ is responsible for the current observational value
of the spectral index $n$, the value of $c$ during inflation is not
determined by $n$. 
In this case, there is an upper bound $c < 0.2$ if there is no fine-tuning
between $\epsilon_H$ and $\eta_\sigma$.}  
Here the definition of the slow-roll parameter $\eta_\sigma$ is
\begin{equation}
\eta_\sigma \equiv \frac{M_p^2 V_{\sigma\sigma}}{V_I}.
\end{equation}
Therefore, the exponential factors in the above equations are at most
$e^{F \Delta N} \simeq e^{0.017 \times 60}<e$. 
Thus, we conclude that the evolution of the curvaton and its fluctuation
during inflation is not significant.

The curvaton dynamics after inflation are different from those of the
inflationary epoch. 
We introduce the barotropic parameter ($w$) of the dominant fluid,
which does not relate to the curvaton, but instead relates 
to the background
radiation ($w=1/3$ for radiation) or the inflaton oscillation (for
inflaton oscillation $w$ depends on the order of the inflaton
potential\cite{turner-coherent}). 
As we are considering the evolution of the curvaton before the beginning
of curvaton oscillation at $m_\sigma \sim H_{osc}$, the dominant
part of the effective potential is
\begin{equation}
V \simeq -\frac{1}{2}(cH)^2\sigma^2.
\end{equation}
Here, both $H$ and $\sigma$ depend on the cosmic time $t$, while $c$ is
a constant determined by the dominant component of the
Universe.
The value of $c$ during inflaton oscillation is believed to be the same
as that of the inflationary epoch, while $c$ may be different during the
radiation-dominated epoch.\footnote{See appendex for more details.}
Substituting the $t$-derivative by $dt\equiv (1+w)d\tau$, we
find\cite{Curvaton_Dynamics} 
\begin{equation}
\sigma'' + (1-w)\sigma' + \frac{4c}{9}\sigma=0,
\end{equation}
where the prime denotes differentiation with respect to $\tau$.
Solving the above equation for $\tau$ and then substituting $t$,
we find\cite{Curvaton_Dynamics}
\begin{equation}
\label{enhg1}
\sigma(t_{out})\simeq \sigma(t_{in})
\left(\frac{t_{out}}{t_{in}}\right)^{-\frac{1}{2}\frac{K}{1+w}},
\end{equation}
where $K$ is defined as
\begin{equation}
\label{enhg2}
K\equiv (1-w)-\sqrt{(1-w)^2 + \frac{16}{9}c^2}.
\end{equation}
Here, $t_{in}$ and $t_{out}$ are the time at the beginning and end of
the evolution, respectively. 
The evolution of the curvaton and its fluctuation is negligible for
$c\simeq 0.2$, while it is significant for $c> 1$.
Since the potential becomes concave near the minimum of the effective
potential, we consider the evolution within
$\sigma(t_{out})<\sigma_{min}$, where $\sigma_{min}$ denotes the
effective minimum for $m_\sigma < cH$.
As the time passes, the bare mass ($m_\sigma$) starts to dominate the
potential and tachyonic amplification stops at
$t=t_{out}$, $(H(t_{out})\simeq m_\sigma)$.
Then, the curvaton oscillation starts at $t=t_{out}$, as for a
traditional curvaton. 

We now consider a concrete example. 
We consider a case in which the coefficient of the induced mass is as
large as $c=5$ after reheating. Reheating occurs at $t=t_{in}$,
then the curvaton oscillation starts at $t=t_{out}=t_{osc}$.
With this simple and generic situation we examine the energy condition
for the hilltop curvaton.
We introduce a new parameter $q$, which gives the reheating temperature
\begin{equation}
T_R \simeq  10^{q/2} m_\sigma.
\end{equation}
Using this parameter, the ratio $t_{out}/t_{in}$ is given by
$t_{out}/t_{in} = 10^{q}$.
Considering Eq.(\ref{enhg1}) and (\ref{enhg2}), we find the enhancement
factor 
\begin{equation}
\gamma\simeq 10^{\frac{1}{2} qc}.
\end{equation}
For $q=2$, the enhancement factor $\gamma$ is given by $\gamma \simeq
10^5$.
Note that the condition $q< -\log (H_I/H_{osc})\equiv -\log (p^2)$ applies
because of the relation
$t_{I} < t_{in}<t_{out}<t_{osc}$, where $t_I$ denotes the time when
inflation ends. 
Since (\ref{bound-alt}) is true for $p=10^{-2}$, we find
\begin{equation}
H_I > 10^{2} {\mathrm GeV}
\end{equation}
for $c=5$, $q=2$ and $p=10^{-2}$.
In this case, we find $\sigma(t_{in}) > 10^{7}$GeV and $\sigma(t_{out})
> 10^{12}$GeV.
The required condition for the effective potential is that
$\sigma(t_{out})$ is smaller than $\sigma_{min}$ during the tachyonic
amplification.  

This significant reduction of the inflation energy is a generic
consequence of the enhancement of the curvaton fluctuation induced by
the $O(H)$ mass. 
The required condition for this amplification are (1) if the coefficient
of the Hubble-induced mass ($c$) is somewhat large during the
radiation-dominated epoch and (2) the sign of the mass 
term is negative, then the curvaton perturbation grows significantly on
a convex potential during the radiation-dominated epoch. 
This significant enhancement gives a significant relaxation of the energy
condition for the primordial inflation.

\section{Conclusions and Discussions}
\hspace*{\parindent}
We have studied a scenario for hilltop curvatons in which a curvaton
evolves on a convex potential. 
We assumed that the effective mass of a curvaton is expressed as 
$m_\sigma = c H$, where the coefficient varies within $0.2 \le c \le 5$
depending on the dominant component of the Universe. 
A negative mass term leads to enhancement of the curvaton fluctuation.
In our model, both $\delta \sigma$ and the ratio $r$ at the
beginning of the oscillation are
enhanced during this epoch, which has a significant impact on the energy
bound for low-scale inflation. 
Following the curvaton hypothesis, we have demonstrated the generality of
this enhancement.  
The Hubble-induced mass is necessary for
the significant enhancement of the curvaton perturbation in our model. 
If there is no $O(H)$ correction to the curvaton potential, our scenario
requires two different kinds of phase transitions that change the
effective mass of the curvaton at least twice. 
These phase transitions must occur (1) between inflation and
tachyonic enhancement and (2) between tachyonic enhancement and curvaton
oscillation.
Moreover, tachyonic enhancement with a constant mass is very dangerous
for our scenario because a curvaton will soon start to oscillate about
the effective minimum.
In order to avoid the reduction of $\delta \sigma$ during this
oscillation, the second phase transition must take place soon after the
first.
The important idea of our scenario is that for a $c \sim O(1)$ potential,
the tachyonic amplification lasts long enough that the conventional
curvaton 
oscillation may start before the beginning of the oscillation about the
effective minimum.

In addition to the condition required for the tachyonic enhancement, we
must consider the origin of the spectral index $n\ne 1$. 
For the curvaton, this index is given by 
\begin{equation}
n-1 \simeq 2\eta_\sigma  -2\epsilon_H,
\end{equation}
where $\epsilon_H\equiv -\dot{H}/H^2$ is the slow-roll parameter.
In many models of inflation in which $\epsilon_H$ is very small,
$\eta_\sigma$ must be responsible for $n\ne 1$.
However, if there is no mechanism that forces 
$|\eta_\sigma|\simeq 0.02$ (i.e., $m_\sigma \sim O(H)$) during
inflation, it is very difficult to attain the present observational
value of the spectral index without introducing fine-tuning.
This may reduce the generality of the curvaton hypothesis, narrowing the
range of possible inflation models in which the curvaton can explain the
curvature perturbation.\footnote{See also Ref.\cite{hilltop-inf}}
In this sense, the $O(H)$ correction is important for the generality of
the curvatons, even when tachyonic enhancement does not play an
important role. 
It should be noted again that the required condition for the curvature
perturbation is very severe for low-scale inflation\cite{low_inflation}
if it is generated by the inflaton fluctuation.

\section{Acknowledgment}
We wish to thank K.Shima for encouragement, and our colleagues at
Tokyo University for their kind hospitality.

\appendix
\section{$O(H)$-corrections}
The purpose of this appendix is to show how $O(H)$ corrections to
light scalar fields appear in the radiation-dominated epoch.
We will mainly follow the original argument given by Dine
et.al, in Ref.\cite{Dine-Rand-Thom}.
The supergravity scalar potential is given by
\begin{equation}
\label{scalar-sugra}
V=e^{\frac{K}{M_p^2}}\left[(D_i W)K^{i\overline{j}}(D_{\overline{j}})
W^* -\frac{3}{M_p^2}|W|^2\right] +V_D,
\end{equation}
where $W$ and $K$ are the superpotential and K\"ahler potential, and the
derivatives are defined as $D_i W=W_i + K_i W /M_p^2$ and $W_i=\partial
W/ \partial \phi_i$.
Looking at this scalar potential, there is a $O(H)$-correction
appears for the soft mass of a light field $\sigma$ when the energy
density is dominated by $F$-terms.
This is the most discussed origin of the
$O(H)$-corrections.
However in the original argument
the origin of such terms is not confined to the scalar potential.
Following Ref.\cite{Dine-Rand-Thom}, we consider $O(H)$
corrections that may appear from nonrenormalizable interactions such as
\begin{equation}
\delta K \propto \Sigma^{\dagger}\Sigma \Phi^\dagger\Phi/M_p^2,
\end{equation}
where the scalar component of the superfield $\Sigma$ is the
light field $\sigma$, and $\Phi$ is a field that dominates the energy
density of the Universe. 
Then, the energy density of the
Universe is given by $\rho\simeq <\int d^4 \theta \Phi^\dagger\Phi>$.
Note that in the thermal phase, the expectation value arises from kinetic
terms, while the usual $\delta m_\sigma^2 \simeq |F_\Phi|^2/M_p^2$
correction is significant in a $F_\Phi$-dominated Universe.
Following this argument, a $O(H)$ correction may result even if D-term
dominates the potential.

From the above argument it is clear that there is no reason to 
assume $c$ in $\delta m_\sigma = cH$ takes exactly the same value in
both the inflation stage and the thermal phase. 
This is the main reason why we considered $O(1)$ variation to $c$.

\end{document}